\documentclass[sigconf, natbib=false]{acmart}
\AtBeginDocument{%
    
}

\RequirePackage[
    datamodel=acmdatamodel,
    style=acmnumeric,
]{biblatex}
\usepackage{subcaption}
\usepackage{algorithm}
\usepackage{algpseudocodex}
\usepackage{tabularx}
\usepackage{multicol}
\usepackage{xspace}
\usepackage{booktabs}
\usepackage{multirow}
\usepackage{subcaption}
\usepackage{soul}
\usepackage{pifont}
\usepackage[normalem]{ulem}
\usepackage{graphicx}
\usepackage{url}
\usepackage{makecell}
\usepackage{nicematrix}
\usepackage{fontawesome}
\usepackage{wasysym}
\usepackage{rotating}
\usepackage{amsthm}

\newcommand{\ashare}[1]{\langle{#1}\rangle}
\newcommand{\bshare}[1]{\langle{#1}\rangle^B}

\begin{document}

\title{SecureV2X: An Efficient and Privacy-Preserving System for Vehicle-to-Everything (V2X) Applications}

\renewcommand{\shorttitle}{SecureV2X}

\author{Joshua Lee}
\email{jlee246@ucsb.edu}
\affiliation{%
    \institution{University of California Santa Barbara}
    \city{Santa Barbara}
    \state{CA}
    \country{USA}
}
\author{Ali Arastehfard}
\email{ali.arastehfard@uconn.edu}
\affiliation{%
    \institution{University of Connecticut}
    \city{Storrs}
    \state{CT}
    \country{USA}
}
\author{Weiran Liu}
\email{weiran.lwr@alibaba-inc.com}
\affiliation{%
    \institution{Alibaba Group}
    \country{China}
}
\author{Xuegang Ban}
\email{banx@uw.edu}
\affiliation{%
    \institution{University of Washington}
    \city{Seattle}
    \state{WA}
    \country{USA}
}
\author{Yuan Hong}
\email{yuan.hong@uconn.edu}
\affiliation{%
    \institution{University of Connecticut}
    \city{Storrs}
    \state{CT}
    \country{USA}
}

\renewcommand{\shortauthors}{J. Lee et al.}

\begin{abstract}
Autonomous driving and V2X technologies have developed rapidly in the past decade, leading to improved safety and efficiency in modern transportation. These systems interact with extensive networks of vehicles, roadside infrastructure, and cloud resources to support their machine learning capabilities. However, the widespread use of machine learning in V2X systems raises issues over the privacy of the data involved. This is particularly concerning for smart-transit and driver safety applications which can implicitly reveal user locations or explicitly disclose medical data such as EEG signals. To resolve these issues, we propose SecureV2X, a scalable, multi-agent system for secure neural network inferences deployed between the server and each vehicle. Under this setting, we study two multi-agent V2X applications: secure drowsiness detection, and secure red-light violation detection. Our system achieves strong performance relative to baselines, and scales efficiently to support a large number of secure computation interactions simultaneously. For instance, SecureV2X is $9.4 \times$ faster, requires $143\times$ fewer computational rounds, and involves $16.6\times$ less communication on drowsiness detection compared to other secure systems. Moreover, it achieves a runtime nearly $100\times$ faster than state-of-the-art benchmarks in object detection tasks for red light violation detection.\footnote{Code is available at \url{https://github.com/datasec-lab/securev2x}.}
\end{abstract}

\keywords{Privacy, Drowsiness Detection, Red-light Violation Detection, Vehicle-to-Everything (V2X)}

\maketitle

\section{Introduction}

Vehicle-to-everything (V2X) technology has played a pivotal role in enhancing road safety, optimizing traffic flow, and enabling autonomous driving technologies.
Recently, the US Department of Transportation (USDOT) released its National Deployment Plan of V2X technologies \cite{usdot_v2x}, calling federal/state/local transportation agencies, the private sector, and researchers to work together to research, test, and deploy V2X technologies and their applications to improve transportation safety, mobility, equity, and sustainability, with ambitious short-term, medium-term, and long-term goals. 
These systems utilize the data generated by various interacting entities (e.g., cars, pedestrians, and cloud servers) to leverage advanced machine/deep learning and computer vision algorithms towards objectives like accident prediction and avoidance, traffic violation detection, and other safety measures. For instance, in the context of crash risk prediction, Shah et al. \cite{shah_novel_2023} employ a multivariate Long Short-Term Memory (LSTM) model to predict the likelihood of a crash between two V2X-enabled vehicles within the next 3 seconds, using data from the previous 7 seconds. It improves predictive capability by incorporating multiple real-time data streams, including cardinal position, acceleration, and speed transmitted by the vehicles.

Similarly, traffic violations such as red light running are often addressed through advanced object detection frameworks, including YOLO, Faster R-CNN, and SSD \cite{bi_achieving_2023, thao_automatic_2022}. These algorithms enable real-time detection and alert mechanisms, ensuring timely interventions. Furthermore, many V2X systems involve infrastructure-to-cloud (I2C) communication, where a third-party server (aka. a mediating agent) handles the computationally intensive deep learning operations required for accurate predictions and detections. This interplay between local and cloud-based computations exemplifies the growing integration of intelligent algorithms in real-world infrastructure, paving the way for safer and more efficient transportation ecosystems. Furthermore, driver drowsiness detection, another safety critical V2X application, also incorporates machine learning on V2X data in a distributed setting. For instance, researchers have used linear regression and convolutional neural networks to identify drowsiness in drivers over a sustained driving task \cite{cui_compact_2022, agarwal_protecting_2019}. When in-vehicle computing resources are limited or the drowsiness detection model is proprietary, inference may be offloaded to a remote edge server. 

Despite their advantages, these inference services require users to disclose 
their private information, such as driving speeds, electroencephalography (EEG) data, user location, license plate details, and other sensitive data points. Additionally, infrastructure that communicates with third-party servers for inference raises similar concerns regarding user data privacy. As a result, the National Deployment Plan of USDOT identified privacy protection as one of their key considerations when testing and deploying V2X technologies and applications \cite{usdot_v2x}. 
Privacy-preserving solutions for V2X techniques require supporting fully functional V2X algorithm computations without leaking users' private data. Secure multi-party computation (MPC)  offers a promising solution to address V2X privacy challenges. Especially in cases where local computation is restricted by the proprietary nature of third-party V2X machine learning algorithms, MPC allows for remote inference while providing user privacy guarantees. Several studies, such as those by Agarwal et al. \cite{agarwal_protecting_2019} and Bi et al. \cite{bi_achieving_2023}, have explored secure computation applications in V2X contexts, including drowsiness detection and object detection for autonomous driving. 
Agarwal et al. \cite{agarwal_protecting_2019} investigate secure distributed training of linear regression across multiple parties for predicting a drowsiness index, while Bi et al. \cite{bi_achieving_2023} and Zhou et al. \cite{zhou2022secure} secure the FasterRCNN and YOLOv3. Despite these advances, the high computational costs of these designs limit their practical, real-time applicability. 

In this work, we focus on introducing advanced and novel MPC techniques to speed up MPC-based V2X applications. We study two representative V2X applications: (1) driver drowsiness detection, and (2) red-light violation detection (based on object detection).
Models integrated for prior privacy-preserving drowsiness detection systems have included linear regression, polynomial regression, and support vector machines (for EEG data analysis and drowsiness detection) \cite{agarwal_protecting_2019, popescu_privacy_2021, liu_privacy-preserving_2022, bi_achieving_2023}. While they have achieved reasonable performance, convolutional networks have demonstrated superior results in drowsiness detection. 

Prior works on privacy-preserving object detection have explored the use of Faster-RCNN and YOLOv3. We follow the advances on object detection and design a MPC-based solution for the YOLOv5 object detector. Previous studies demonstrate that YOLOv5 achieves inference speeds approximately 10 times faster than Faster-RCNN, which is crucial for latency-sensitive applications, and may also brings efficiency improvement in the MPC setting. Furthermore, YOLOv5 maintains sufficient average bounding box precision and object classification accuracy for a range of use cases \cite{yusro_mod_comparison_2022, ahmed_pothole_detection_2021, khalfaoui_compare_2022, nepal2022comparing}. It also offers enhanced flexibility for practitioners by providing multiple model sizes, allowing for the tuning of accuracy and inference speed to suit specific tasks. 

Ensuring privacy for YOLOv5 presents additional challenges. Specifically, designing MPC solutions for all five versions of the YOLOv5 model 
requires the design of a unique and extremely detailed protocol for each. Moreover, YOLOv5's use of swish (\textit{sigmoid linear unit}) activations -- which help it achieve better accuracy and precision in the plaintext setting \cite{zhang2023fine} -- requires more complex design compared to LeakyReLU which was used in YOLOv3 \cite{zhou2022secure}, and ReLU in Faster-RCNN \cite{bi_achieving_2023}.

Existing MPC-based machine learning libraries provide secure protocols for a wide range of functions, but they cannot perform inference on models for V2X applications. In particular, frameworks such as Delphi, Gazelle, and CrypTFlow2 
provide only the most fundamental primitives, such as secure 
convolutions, fully connected layers, argmax, maxpooling, and 
ReLU. While CrypTen offers a broader range of secure functionalities, it remains limited in flexibility. This constraint prevents most machine learning practitioners from seamlessly integrating secure computation into their workflows. To address this gap, we propose novel secure functionalities for missing operations including secure upsampling, split, and constant floor. Additionally architectural modifications are made to ensure compatibility with secure drowsiness detection.

Moreover, we revise CrypTen's underlying network graph processor to 
support empty-argument ONNX functions. This This drastically increases the number of V2X and machine learning applications for which downstream application is immediately accessible. 
Although this paper focuses on driver drowsiness and red light violation detection, the enhanced flexibility of SecureV2X compared to prior works ensures that it can easily be extended to problems such as real-time traffic flow analysis, accident prediction, and pedestrian safety. Furthermore, the behavior and complexity of operations in plaintext settings differ significantly from those in secure computation, making the selection of appropriate secure primitives and parameters for our models a complex task. We address this challenge through extensive analysis of secure primitives and hyperparameters, evaluating their impact on overall performance.

Thus, our contributions are summarized as follows: 

\begin{itemize}
    \item To our best knowledge, we propose the first real-time privacy preserving system SecureV2X for two practical V2X applications: (1) EEG-based driver drowsiness detection, and (2) red-light violation detection via YOLOv5. 

    \item We design and implement the privacy-preserving system SecureV2X with novel cryptographic protocol constructions, which offers provable security. 
 
    \item SecureV2X significantly outperforms the state-of-the-art systems for privately detecting drowsiness and red-light violation in intelligent transportation systems.

\end{itemize}

\section{SecureV2X Framework}

\subsection{V2X Applications}

\noindent\textbf{Drowsiness Detection}. 
Various methods exist for computing drowsiness detection \cite{ramzan2019survey}. The state-of-the-art solution relies specifically on single-channel EEG data which has been shown to be highly predictive in distinguishing between drowsy and alert driving states, as discussed in \cite{cui_compact_2022}. 

To perform drowsiness detections over an input EEG signal, we adopt the CompactCNN model \cite{cui_compact_2022} and its corresponding data preprocessing approach. Specifically, preprocessed EEG signals (filtered by a 1-Hz high-pass and 50-Hz low pass finite impulse response filter with muscular and ocular artifacts removed from the signal) are down-sampled from 500 Hz to 128 Hz and extracted as 3-second samples. A single channel of the EEG sample is then selected, resulting in an input vector $x \in R^{1 \times 384}$. CompactCNN utilizes data sourced from a series of multi-channel EEG recordings during a ``sustained-attention driving task'' and uses local reaction time (the time between when a car would begin to drift, and when the driver would respond) and global reaction time (the average reaction time over trials within a 90-second window before the onset of the deviation) to measure levels of instantaneous drowsiness, accounting for baseline differences in driver reaction time \cite{lin2005eeg}. If reaction times were greater than 2.5 times the baseline, the driver was considered to be drowsy, whereas reaction times shorter than 1.5 times the baseline were labeled as ``alert''.  

CompactCNN consists of a shallow network architecture. Namely, a single convolution layer of 32 filters of size 1 x 64 is followed by batch normalization which helps to minimize covariate shift and accelerate training convergence \cite{cui_compact_2022}. The ELU activation function is applied over this output to capture non-linear patterns in the data. These operations generate 32 $(1\times 321)$ feature maps:

\vspace{-0.15in}

\begin{align*}
\small
& \text{Conv}(\mathcal{O}_{i},W_{i,\ell},b_{i,\ell})_{k} = \sum_{j=1}^{64}x_{k+j}\cdot W_{i,j,\ell}, k\in[1,n]\\
& \mathcal{O}_{i+1} = \text{ELU}(\text{Batch}(\text{Conv}(\mathcal{O}_{i}, W_{i,\ell}, b_{i,\ell})_{1\times n})) \\
& \hspace{0.75cm} = \text{ELU}\left(\gamma\frac{\text{Conv}(\cdot)_{1\times n} - \mu}{\sigma} + \beta\frac{\text{Conv}(\cdot)_{1\times n} - \mu}{\sigma}\right)
\end{align*}

Here, $\text{ELU}$ represents the exponential linear unit, $\gamma$ and $\beta$ indicate batch normalization weights and biases, $\mu$ and $\sigma$ are the aggregated mean and standard deviation for the feature map output by $\text{Conv}(\cdot)$ over the training data, and $\mathcal{O}_{i}, W_{i,\ell}$ and $b_{i,\ell}$ are the input tensor and filter weights and biases (for the $\ell$-th filter) of the $i$-th layer.
Global average pooling is conducted on the activation output, and a dense layer generates the final output from the pooled feature vector (1 x 32) to obtain an output of shape 1 x 2. This is passed through the log-softmax function to obtain a binary classification with regards to drowsiness. So for our final detection, we obtain 

\vspace{-0.1in}

\begin{equation}
\mathcal{O}_{i+2} = \text{GAP}(\mathcal{O}_{i+1}) \cdot W_{i+1}
\end{equation}
\begin{equation}
detection = \log\left(\frac{e^{\mathcal{O}_{i+2}[1]}}{\sum_{s=0}^{1}e^{\mathcal{O}_{i+2}[s]}}\right)
\end{equation}

\begin{figure*}[htbp]
\centering
\includegraphics[width=\textwidth]{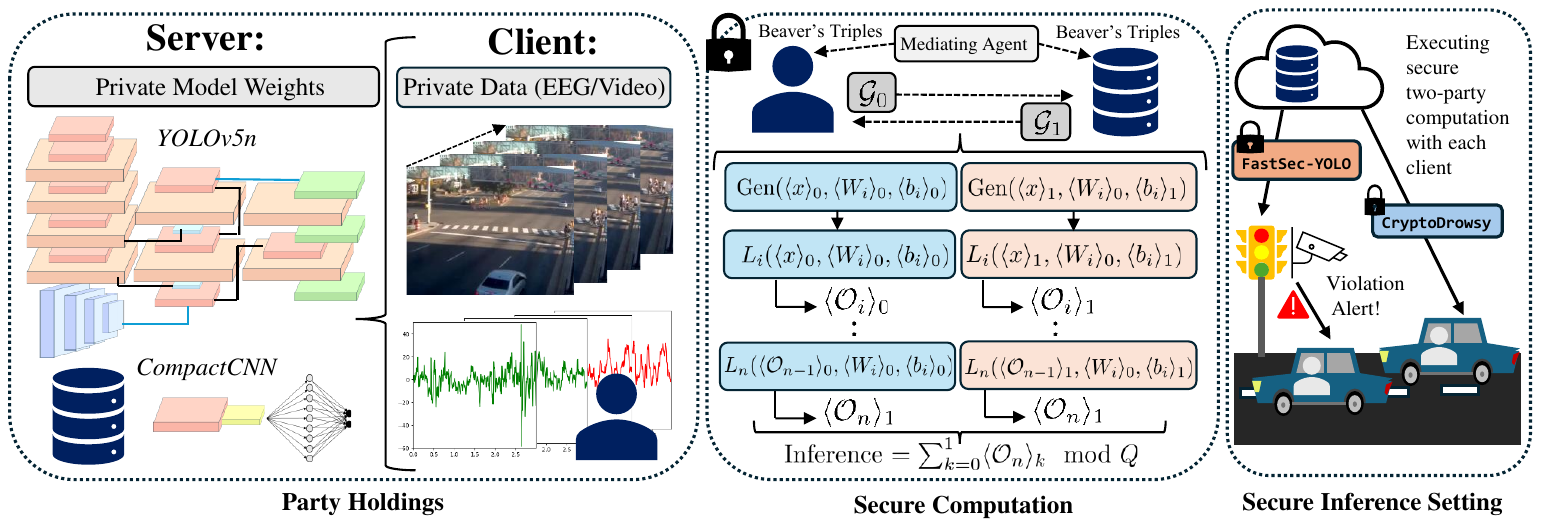}\vspace{-0.1in}
\caption{Overview for the SecureV2X Framework. Server: edge server or the cloud (infrastructure), Client: connected vehicle and/or traffic camera. The server is executing secure two-party computation with each client.}\vspace{-0.15in}
\label{fig:system_design}
\end{figure*}

\noindent\textbf{Red-Light Violation Detection via YOLOv5}. 
Red-light violation detection is a critical area of research in traffic management. When vehicles fail to comply with traffic signals, the consequences can include property damage, injuries, and fatalities \cite{cohn2020red, aaa2019red,studdert2017once}. Additionally, red light violation detection serves as a valuable application of V2X data. 
To detect red light violations from video feed data, we utilize the detection algorithm proposed by \cite{yahya2019fully}. First, standard image processing techniques including gray-scale binary thresholding, dilation and erosion are used to identify the detection line. A series of contours are generated and the noisy contours are filtered out. This leaves behind the crosswalk's contour components. These contours are used to generate the violation detection line. Vehicles detected beyond this point in the video feed are determined to be violators.

Second, an initial object detection is performed over the first frame. Each of these objects is assigned
to a median Flow tracker \cite{kalal2010forward} and tracker positions are updated every frame. For every fifth frame, an object detector scans the input to identify new vehicles in the scene and improve the precision of tracking coordinates for vehicles which are already have a tracker. These object detections are conducted by the YOLO algorithm. YOLO is a widely used machine learning algorithm in computer vision, optimized for efficiently detecting bounding boxes for objects in images. 

In the original approach by Yahya et al. \cite{yahya2019fully}, YOLOv3 was used for this purpose. However, our system leverages the YOLOv5 model by Ultralytics. YOLOv5 comprises three main components: the backbone, neck, and detection head. The backbone extracts low-level image features, while the neck aggregates information from various depths of the backbone, forming a hierarchical feature structure that enhances detection accuracy. The detection head then generates the final bounding box predictions.

In addition to object tracking and detection, a color histogram is used to determine the color of the traffic light at any given frame. If the color is red, and a tracked vehicle's center coordinate is determined to be beyond the violation line at that point, then it is considered a red light violation. 

\subsection{SecureV2X Framework}\label{sec:secv2x}

Figure \ref{fig:system_design} shows that SecureV2X facilitates the secure and efficient computation of drowsiness and red-light violation detection. An edge server hosts proprietary weights for the YOLOv5n and CompactCNN models, while a connected vehicle or traffic camera (user/client) provides EEG signal data or traffic video data. Both parties engage in a secure computation protocol to jointly generate the desired output.

CryptoDrowsy performs drowsiness detection by establishing a secure communication channel between the vehicle and the edge server hosting the model. Secret shares are generated for the vehicle's EEG signal and the server's model weights. A secure mediating agent then distributes Beaver's triples for multiplications, enabling inference. Once inference is complete, the server sends its share of the output to the vehicle, which reconstructs the result. Further details on this can be found in Section \ref{sec:crypten}.

For red-light violation detection, vehicles at or approaching an intersection connect to a traffic camera. The camera runs the detection algorithm locally, except for object detection which requires a secure connection to a model-hosting edge server. Using a secure mediating agent, the server and camera exchange random multiplicative triples to jointly compute bounding boxes over video frames. After computation, the server sends its share of the output to the camera, which reconstructs the results and applies non-maximum suppression to refine bounding boxes. If a violation is detected, the camera alerts the offending vehicle. 
A visual overview of this system can be seen in Figure \ref{fig:system_design}.

\section{Crypto-Drowsy}




\begin{algorithm}
\small
    \caption{CryptoDrowsy}
    \label{alg:crypto_drowsy}
    \begin{algorithmic}[1]
    \Statex \textbf{Input:} $P_{0}$ holds EEG sample $x\in\mathbb{R}^{1\times384}$ and $P_{1}$
    holds model weights $W=\{W_{1},W_{2},W_{3}\}$ and biases $b=\{b_{1},b_{2},b_{3}\}$
    \Statex \textbf{Output:} $P_{0}$ learns drowsiness classification, $c \in \{0,1\}$
    \State $P_{k}$ for parties $k \in \{0,1\}$ invoke $\mathcal{F}_{PRZS}$ to generate secret shares of their inputs, $\langle x \rangle_{k}, \langle W_{i} \rangle_{k}, \langle b_{i} \rangle_{k}$, where $i \in \{1, 2, 3\}$
    \BeginBox[draw=blue]
        \State $P_{k}$ for parties $k \in \{0,1\}$ invoke $\mathcal{F}_{Conv2D}$ and 
        provide input shares $\langle x \rangle_{k}, \langle W_{1}\rangle_{k}, 
        \langle b_{1} \rangle_{k}$ and obtain $\langle\mathcal{O}_{1}\rangle_{k}$
    \EndBox
    \BeginBox[draw=cyan]
        \State $P_{k}$ for parties $k \in \{0,1\}$ call $\mathcal{F}_{BN}$ on shares 
        $\langle\mathcal{O}_{1}\rangle_{k}, \langle W_{2}\rangle_{k}, \langle b_{2}\rangle_{k}$
        to obtain $\langle\mathcal{O}_{2}\rangle_{k}$
        \State $P_{k}$ for parties $k \in \{0,1\}$ call $\mathcal{F}_{ReLU}$ and input shares 
        $\langle\mathcal{O}_{2}\rangle_{k}$ and learn $\langle\mathcal{O}'_2\rangle_{k}$ 
        \State $P_{k}$ for parties $k \in \{0,1\}$ reshape shares $\langle\mathcal{O}'_2\rangle_{k}$
        to $1 \times 32 \times 321$ and perform global pooling for each $1 \times 321$ feature 
        map to output $\langle\mathcal{O}''_{2}\rangle_{k}$ of shape $1 \times 32$
    \EndBox
    \BeginBox[draw=teal]
        \State $P_{k}$ for parties $k \in \{0,1\}$ input $\langle\mathcal{O}''_{2}\rangle_{k}$, $\ashare{W_{3}}_{k}$ and $\ashare{b_{3}}_{k}$
        to $\mathcal{F}_{FC}$ and output $\langle\mathcal{O}_{3}\rangle_{k}$
    \EndBox 
    \State $P_{1}$ sends its output share $\langle\mathcal{O}_{3}\rangle_1$ to $P_{0}$ who 
    reconstructs $c = \ashare{O_3}_0 + \ashare{O_3}_1$
    \end{algorithmic}
\end{algorithm}

\subsection{Secure Construction for the System}
\label{sec:crypten}

The problem of privately detecting driver drowsiness over a user (driver - Party $P_0$) and the service/model provider (Party $P_1$) is formulated as follows. $P_0$ holds an EEG sample vector $x \in R^{m \times n}$, and $P_1$ holds weights $W = \{w_{1}, w_{2}, \dots, w_{n}\}$ for the EEG signal classification model. $P_0$ and $P_1$ jointly compute the classification of $x$, denoted as $c \in \{0,1\}$, where $0$ indicates a negative classification (the driver is not drowsy) and $1$ indicates a positive classification (the driver is drowsy). During the classification procedure, we require that $P_0$ learns nothing about $P_1$'s model (the proprietary pretrained weights of a publicly known architecture) except for its inference output, and $P_1$ obtains no information about $P_0$'s input data and its inference output.

To maintain the privacy of pretrained model weights (held by the service provider) and input data (held by the user), we design a specific secure protocol for driver drowsiness detection, and utilize the CrypTen MPC framework \cite{knott_crypten_2022} for the concrete implementation. 

Algorithm \ref{alg:crypto_drowsy} shows the detailed protocol description. Here, we show its design idea. Under the CrypTen framework, $P_0$ or $P_1$ holds secret shares of the other party's inputs (the EEG sample and model weights owned by the user and the service provider, respectively). To share an input $x$, a pseudorandom zero-share (PRZS) \cite{cramer2005share} is first generated by $P_0$ and $P_1$ that sum to $0$. Then, the party that holds the input $x$ adds $x$ to their share and discards $x$. For ease of description, in the rest of the paper, we denote $\ashare{x}$ as the shares of $x$, and $\ashare{x}_0, \ashare{x}_1$ as the shares of $x$ owned by $P_0$ and $P_1$, respectively.
All computations $f$ involved for the inference done by CrypTen follows the paradigm that $P_0$ and $P_1$ own $\ashare{f(x)}$. Therefore, it is easy to reveal the final inference result, that is, the service provider $P_1$ sends its output share to the user $P_0$ so that the user obtains its inference result as the sum of the resulting shares. 

Additive shares have homomorphic properties, which allow for private addition, multiplication, comparison, and other non-linear computations. Given two shares $\ashare{x}$, $\ashare{y}$, private addition can be easily done by having $P_0$ and $P_1$ compute $\ashare{z}_i = \ashare{x}_i + \ashare{y}_i$ for $i \in \{0, 1\}$, respectively. Multiplication of a public value $p$ and a share $\ashare{x}$ can be also easily done by having $P_0$ and $P_1$ compute $\ashare{z}_i = p \cdot \ashare{x}_i$.
Private multiplication is non-trivial. CrypTen introduces a helper party $T$ to assist with private multiplication by generating Beaver triples \cite{beaver1991efficient} $(\ashare{a}, \ashare{b}, \ashare{c})$ where $c = ab$.
With the help of the Beaver triple, $P_0$ and $P_1$ compute $\ashare{e}  = \ashare{x} - \ashare{a}$ and $\ashare{d} = \ashare{y} - \ashare{b}$, and reveal the output to obtain public values $e$ and $d$. This enables the computation of $\ashare{x}\ashare{y} = \ashare{c} + e\ashare{b} + \ashare{a}d + ed$. 
Private comparison between $\ashare{x}$ and $\ashare{y}$  extracts the highest shared bit of the subtraction of $\ashare{x-y}$, which can be efficiently done via a special MPC protocol B2A~\cite{demmler2015aby}. Additional details about these conversions are provided in Appendix \ref{sec_proofs_conversion}.

We now detail the computations involved in CompactCNN. Input shares are first generated, and Beaver triples are distributed by the helper to compute the initial convolution. Batch normalization is applied to the resulting arithmetic shares using CrypTen’s summation, public division, and variance functions. ReLU activation employs a multiplexer to compute $\ashare{x > 0}$, setting $\ashare{1}$ if $x > 0$ or $\ashare{0}$ otherwise. ReLU is then calculated as $\ashare{x}\ashare{x > 0}$.

Global average pooling computes the mean across the $1 \times 321$ feature map by summing the secret-shared tensor and performing public division to avoid fixed-point encoding errors (see CrypTen~\cite{knott_crypten_2022}). The dense layer leverages Beaver triples for secure matrix multiplication between the weights tensor and the input. The network's output is finalized using CrypTen’s approximations for the log-softmax function, applied directly to secret-shared data via Newton-Raphson iterations for exponentiation and maximum/summation protocols.

All functions rely on CrypTen’s secure multiparty computation primitives. Once computation is complete, the server sends its share of the result to the user, who combines it with their own and scales down to reconstruct the plaintext output. The security of these operations within CryptoDrowsy is discussed in section \ref{sec_proof_secv2x}.

\section{FastSec-YOLO}

\subsection{Secure Construction for the System}

As explained in section \ref{sec:secv2x}, we do not implement the entire RLR detection algorithm in a private context. Instead, we leverage CrypTen to convert YOLOv5 to a CrypTensor-based network, enabling the generation of secret shares for both the provider's model and the user's data. In this setup, most components of the algorithm are computed locally, with only object detection performed via secure two-party computation.

CrypTen seamlessly converts most PyTorch models into its framework, thanks to its extensive library of secure functions. However, key components of the YOLOv5 architecture, such as upsampling and tensor splitting, were not originally supported. To address this, we developed customized secure protocols for these operations, enabling the use of YOLOv5 for secure object detection. A detailed description of the protocol can be found in Algorithm \ref{alg:fastsec-yolo}.

As with CryptoDrowsy, FastSec-YOLO enables a client (Party $P_{0}$) 
and a model provider (Party $P_{1}$) over secretly shared input data and model
weights owned by $P_{0}$ and $P_{1}$, respectively. Here, $P_{0}$ holds a square image $x\in\mathbb{R}^{m\times m\times 3}$ where $m$ is some multiple of 
$32$, and $P_{1}$ holds weights $W = \{w_{1},w_{2},\dots,w_{n}\}$ for the trained YOLOv5n object detector. Together, $P_{0}$ and $P_{1}$ 
compute a detection output of shape $\mathcal{O}\in\mathbb{R}^{n\times 85}$
where $n$ is the number of detections made, and the first five columns correspond to the object class, $x$ and $y$ center values, the width, and the height of the output bounding box, respectively. The other 80 columns indicate
the confidence estimate for each possible detection class. 

Here, we present the protocol design for FastSec-YOLO. Additional details  about its construction can be seen in figure \ref{alg:fastsec-yolo}. 
As with CryptoDrowsy, computation starts with the generation of 
secret shares through the PRZS algorithm, and the distribution of 
Beaver's triples involved in the convolutions performed over YOLOv5n. The semantics of these procedures are identical to those involved in computing CryptoDrowsy and can be referred to in Section \ref{sec:crypten}.

FastSec-YOLO's computation involves three core components: the backbone, the neck, and the detection head. First, the CSPDarknet backbone of YOLOv5n computes feature maps from the input image, with cross-stage connections facilitating the flow of information from lower network levels to the feature aggregation neck. Next, the PANet neck aggregates these features and passes them to the detection head, which outputs bounding box detections, classifications, and confidence estimates for each class. Each network component utilizes C3 convolution blocks, incorporating convolution, batch normalization, and Swish activation functions. The secure protocol for computing these blocks is detailed in Algorithm \ref{alg:c3} in the appendix. At the backbone's end, spatial pyramid pooling extracts the network's most critical features, with the secure implementation described in Algorithm \ref{alg:sppf} in the Appendix. Finally, the aggregation neck applies secure upsampling and further convolution operations to enhance bounding box detection accuracy, particularly for small objects.

 \begin{algorithm}
\small
    \caption{FastSec-YOLO}
    \label{alg:fastsec-yolo}
    \begin{algorithmic}[1]
        \Statex \textbf{Input:} $P_{0}$ holds normalized image $x\in\mathbb{R}^{k\times l\times 3}$
        and $P_{1}$ holds weight sets $W=\{W_{1},W_{2},\dots,W_{n}\}$ and bias sets 
        $b=\{b_{1},b_{2},\dots,b_{n}\}$
        \Statex \textbf{Output:} $P_{0}$ learns output $z\in\mathbb{R}^{m\times85}$ where $m$ is
        the number of raw bounding boxes generated
        \State Parties $P_{k},k\in\{0,1\}$ invoke $\mathcal{F}_{PRZS}$ to obtain secret shares of $x, W$, and $b$, $\langle x\rangle_{k}, \langle W_{i}\rangle_{k}, \langle b_{i}\rangle_{k}$
        for all $i\in\{1, 2,\dots,n\}$
        \State Initialize dictionary \textit{ResWeights}
        \State For $k \in \{0, 1\}$, $\ashare{\mathcal{O}_0}_k \leftarrow \ashare{x}_k$
        \For{$i=1, 2,\dots,n$}
            \If {$L_{i} = $ ConvBNSiLU}
                \State $\langle\mathcal{O}_{i}\rangle_{k} \leftarrow 
                \mathcal{F}_{ConvBNSiLU}(\langle\mathcal{O}_{i-1}\rangle_{k},
                    \langle W_{i}\rangle_{k}, \langle b_{i}\rangle_{k})$
                \If {ConvBNSiLU has residual connection}
                    \State \textit{ResWeights}[i] = $\langle\mathcal{O}_{i}\rangle_{k}$
                \EndIf
            \ElsIf {$L_{i} = $ Conv2D}
                \State $\langle\mathcal{O}_{i}\rangle_{k} \leftarrow 
                    \mathcal{F}_{Conv2D}(\langle\mathcal{O}_{i-1}\rangle_{k},
                    \langle W_{i}\rangle_{k}, \langle b_{i}\rangle_{k})$
                \If {Conv2D has residual connection} 
                    \State \textit{ResWeights}[i] = 
                    $\langle\mathcal{O}_{i}\rangle_{k}$
                \EndIf
            \ElsIf {$L_{i} = $ C3}
                \State $\langle\mathcal{O}_{i}\rangle_{k} \leftarrow 
                    \mathcal{F}_{C3}(\langle\mathcal{O}_{i-1}\rangle_{k},
                    \langle W_{i}\rangle_{k}, \langle b_{i}\rangle_{k})$
                \If {C3 has residual connection} 
                    \State \textit{ResWeights}[i] = 
                    $\langle\mathcal{O}_{i}\rangle_{k}$
                \EndIf
            \ElsIf {$L_{i} = $ SPPF}
                \State $\langle\mathcal{O}_{i}\rangle_{k} \leftarrow 
                    \mathcal{F}_{SPPF}(\langle\mathcal{O}_{i-1}\rangle_{k},
                    \langle W_{i}\rangle_{k}, \langle b_{i}\rangle_{k})$
            \ElsIf {$L_{i} = $ Concat $(\ell)$}
                \State $\langle\mathcal{O}_{i}\rangle_{k} \leftarrow 
                    \mathcal{F}_{Concat}(\langle\mathcal{O}_{i-1}\rangle_{k},$
                    \textit{ResWeights}[$\ell$]$)$
            \ElsIf {$L_{i} = $ Upsample}
                \State $\langle\mathcal{O}_{i}\rangle_{k} \leftarrow 
                    \mathcal{F}_{Upsample}(\langle\mathcal{O}_{i-1}\rangle_{k})$
            \EndIf
        \EndFor
    \State $P_{1}$ sends its share $\langle\mathcal{O}_{n}\rangle_{k}$ to 
        $P_{0}$ who reconstructs the output $z$ from its shares
    \end{algorithmic}
\end{algorithm}

After the forward pass of FastSec-YOLO, the model holder sends its output shares to the data holder. In this scheme, both parties must agree on how the shares are redistributed, ensuring that the server cannot access the user's output data— which is inherently sensitive—without permission. Under the semi-honest model assumption, the protocol dictates that the server sends its shares to the user, while the user does not return its own shares to the server. The user then sums and scales its shares to reconstruct the final output bounding boxes.
The underlying security of SecureV2X -- and by extension FastSec-YOLO and CryptoDrowsy -- is discussed further in section \ref{sec_proof_secv2x}.

\section{Security of SecureV2X}\label{sec_proof_secv2x}

SecureV2X is composed of CryptoDrowsy and FastSec-YOLO, each implementing functionalities for computing convolutions, batch normalization, max pooling, sigmoid linear unit activations, ReLU activations, global pooling, fully connected layers, concatenation, and upsampling. These functionalities rely on combinations of primitive operations, including addition, multiplication, comparison, and public division. Non-linear functions, such as the inverse square root (used in batch normalization) and the sigmoid, are implemented using linear approximations based on these primitive operations.

Both CryptoDrowsy and FastSec-YOLO rely on the secure primitives provided by CrypTen \cite{knott_crypten_2022}. All secure functions implemented by CrypTen involve arithmetic and binary secret-sharing, and conversions between the two for efficiency. As demonstrated in \cite{damgard2012multiparty}, the security of functions is preserved under composition. Therefore, the security of our system and all involved functions is derived from the security of the compositions of these secure primitives.

Tensor operations such as 
concatenation, upsampling, and splitting are non-interactive, meaning they can be performed locally on secret-shared data. Thus, their security trivially follows from the security of the secret-sharing paradigm in use. Addition operations performed over arithmetic secret shares are also non-interactive, which means that they are trivially secure as well. 
All secure multiplication and addition operations are performed over arithmetic secret shares in the ring $\mathbb{Z}_{2^{L}}$, which have been proven secure in \cite{bogdanov2008sharemind, knott_crypten_2022}.
 
\subsection{Secure Operations over Binary Secret Shares}\label{sec_proofs_comparison}

Secure comparison underlies the secure activation functions provided by the CrypTen framework. These operations are performed over binary secret shares, which are more efficient for this task. Additionally, logic and multiplexing operations involved in max pooling, an essential component of the spatial pyramid pooling block in YOLOv5's architecture, are performed over binary secret shares. Operations conducted over binary secret shares are represented as Boolean circuits using the GMW protocol \cite{goldreich1987how}. Three fundamental logic operators are defined in this setting.

First, \texttt{XOR} can be computed locally on each party's secret shares. Suppose $\bshare{z} = \bshare{x} \oplus \bshare{y}$. Then, $P_{i}$ locally computes $\bshare{z}_{i} = \bshare{x}_{i} \oplus \bshare{y}_{i}$. Since this protocol involves no communication, the operation is non-interactive and secure.

Second, \texttt{AND} operations are computed over these shares. For $\bshare{z} = \bshare{x} \land \bshare{y}$, a precomputed Boolean triple $\bshare{c} = \bshare{a} \land \bshare{b}$, where $a$ and $b$ consist of bits sampled uniformly at random, is used to compute the output since bitwise \texttt{AND} is equivalent to multiplication modulo 2. As described in \cite{demmler2015aby}, $P_{i}$ obtains $\bshare{e} = \bshare{a}_{i} \oplus \bshare{x}_{i}$ and $\bshare{f}_{i} = \bshare{b}_{i} \oplus \bshare{y}_{i}$. Each party then reconstructs $e$ and $f$ by exchanging their respective shares, $\bshare{e}_{i}$ and $\bshare{f}_{i}$, and computing the \texttt{XOR} of the received shares. Since $a$ and $b$ are sampled at random, the values of $e$ and $f$ reveal no information about $x$ or $y$. Finally, $P_{i}$ computes $i \cdot e \cdot f \oplus f \cdot \bshare{a}_{i} \oplus e \cdot \bshare{b}_{i} \oplus \bshare{c}_{i} = \bshare{z}_{i}$.

Lastly, \texttt{Bit-Shift} operations can be efficiently computed over binary secret shares. When parties wish to shift the bits of a binary secret-shared value $\bshare{x}$ by a constant $k$, each party locally computes $\bshare{y}_{k} = \bshare{x}_{k} >> k$. Since each bit of a binary secret-shared value is independent of all other secret-shared bits, bit shifts can be easily performed over binary secret shares. The non-interactivity of this operation ensures its security. 

\subsection{A2B and B2A Secret Share Conversions}\label{sec_proofs_conversion}

Because some operations are more efficiently computed over arithmetic secret shares and others over binary secret shares, it is important to use efficient conversions between the two types. CrypTen's primitives for computing max pooling, ReLU, and other functionalities convert input arithmetic secret shares to binary secret shares, compute the logical function, and then convert the output back into arithmetic secret shares.

The security of arithmetic-to-binary (A2B) secret share conversions is demonstrated in \cite{damgard2006unconditionally}. Each party secretly transmits its arithmetic share to the other and then performs addition over both shares. Next, both parties instantiate binary secret shares $\bshare{y}_{k}$, where each $y_{k}$ is an arithmetic secret share, i.e., $y_{k} = \ashare{x}_{k}$. Once these secret shares are defined, each party computes $\ashare{x} = \sum_{k \in \mathcal{P}} \bshare{y}_{k}$. CrypTen's primitives employ two approaches to compute this sum. If memory is sufficient, a carry-lookahead adder circuit is used to evaluate the sum in $\log_{2}(|\mathcal{P}|)\log_{2}(L)$ rounds. However, if GPU memory is exceeded, a less memory-demanding protocol is adopted, which requires $|\mathcal{P}|\log_{2}(L)$ rounds to compute. 

Converting binary shares to arithmetic shares (B2A) is accomplished as follows. First, parties compute $\ashare{x} = \sum_{b=1}^{B} 2^{b} \ashare{\ashare{x}^{(b)}}$, where $\ashare{x}^{(b)}$ indicates the $b$-th bit of the binary share $\bshare{x}$ ($B$ is the total number of bits in the secret being shared). The arithmetic shares of bits are created using $B$ pairs of secret-shared bits $(\ashare{r}, \bshare{r})$ during the offline phase. Each pair represents the bit-value $r$ in both arithmetic and binary secret-share forms. Next, each party computes $\bshare{z} \leftarrow \bshare{b} \oplus \bshare{r}$. This result is revealed and used to compute $\ashare{b} \leftarrow \ashare{r} + z - 2\ashare{r}z$. In this scheme, the secret shares $\bshare{r}$ are chosen uniformly at random. Consequently, the revealed value $\bshare{z} \leftarrow \bshare{b} \oplus \bshare{r}$ is \textit{indistinguishable from white Bernoulli random noise}. This ensures the security of the conversion. For more detailed proof of this, see \S C.1.3 in \cite{knott_crypten_2022}.

\section{Experimental Results}

\begin{figure*}[!htbp]
    \centering
    \begin{subfigure}[b]{0.25\textwidth}
        \centering
        \includegraphics[width=\textwidth]{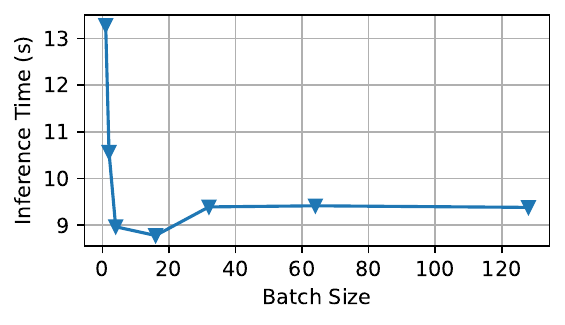}
        \caption{Secure Inf. by Batch Size}
        \label{graph:sec_batch}
    \end{subfigure}
    \hspace{-0.1in}
    \begin{subfigure}[b]{0.25\textwidth}
        \centering
        \includegraphics[width=\textwidth]{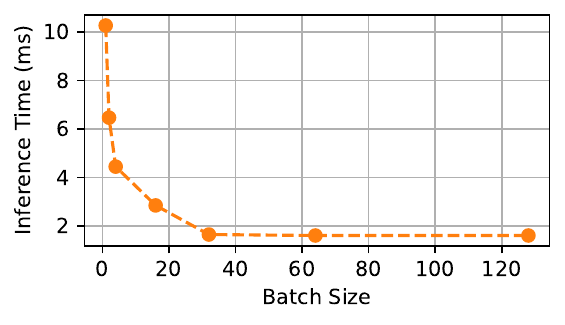}
        \caption{Plain Inf. by Batch Size}
        \label{graph:plain_batch}
    \end{subfigure}
    \hspace{-0.1in}
    \begin{subfigure}[b]{0.25\textwidth}
        \centering
        \includegraphics[width=\textwidth]{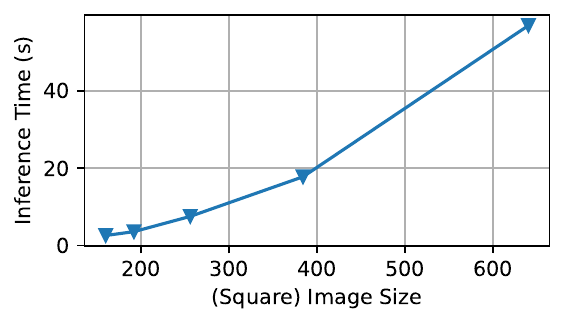}
        \caption{Secure Inf. by Image Size}
        \label{graph:sec_image}
    \end{subfigure}
    \hspace{-0.1in}
    \begin{subfigure}[b]{0.25\textwidth}
        \centering
        \includegraphics[width=\textwidth]{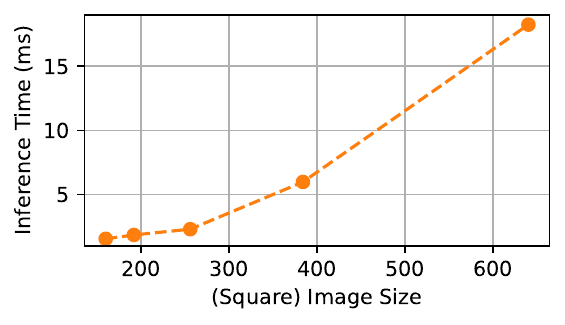}
        \caption{Plain Inf. by Image Size}
        \label{graph:plain_img}
    \end{subfigure}
     \caption{Evaluation of FastSec-YOLO on Varying Batch and Image Sizes under the Secure and Plaintext Settings. We compute object detection for batch sizes of 1, 2, 4, 16, 32, 64 and image resolutions of $160\times 160$, $192\times 192$, $256\times 256$, $384\times 384$ and $640\times 640$.}
    \label{fig:parameter_exps}
\end{figure*}

\begin{table*}[!htbp]
\caption{MAP50, Mean Precision (MP), Mean Recall (MR), Communication Cost, and Runtime (efficiency) metrics for each private object detection system over vehicle classes. All systems, have the same fixed parameters (batch size is 32, image size is 288 $\times$ 288).}
\vspace{-0.1in}
\centering
\begin{tabular}{|c|c|c|c|c|c|c|} 
\hline 
System & MAP50 & MR & CPU Time (s) & GPU Time (s) & Com. (MB) & Rounds \\
\hline 
Fast Sec-YOLO & 0.4388 & 0.3425 & 8.7781 & 1.734 & 151.5 & 77.5 \\
CrypTen-YOLOv5s & 0.4563 & 0.3678 & 19.2428 & 3.593 & 288.0 & 77.5 \\
CrypTen-YOLOv5m & 0.6162 & 0.5241 & 36.9849 & 8.000 & 497.2 & 102.5 \\
CrypTen-YOLOv5l & 0.6918 & 0.6136 & 59.8448 & 14.292 & 755 & 127.5 \\
CrypTen-YOLOv5x & 0.7549 & 0.6460 & 97.0974 & 26.017 & 1,061.5 & 152.5 \\
\hline
\end{tabular}
\label{tab:fast_sec_yolo}
\end{table*}

For both CryptoDrowsy and FastSec-YOLO, we evaluate performance using three key metrics: accuracy, communication, and efficiency. Secure computations rely on fixed-point representations instead of high-precision floating-point values used in plaintext settings, leading to minor performance degradations in privacy-preserving machine learning. Additionally, certain secure operations require approximations for efficiency, further impacting accuracy. Measuring the extent of these degradations across different secure systems is crucial. The main obstacle to the widespread adoption of secure computation protocols in machine learning and V2X settings is their inefficiency compared to plaintext counterparts. Therefore, our evaluation primarily focuses on how efficiently these computations can be performed under various parameters and configurations.

\subsection{Secure Drowsiness Detection}

\begin{table}[!htbp]
\small
\caption{Average runtime, communication, and round complexity performance on private drowsiness detection (our CryptoDrowsy vs. relevant benchmarks for a batch of 314 EEG signals).}
\vspace{-0.1in}
\centering
\begin{tabular}{|c|c|c|c|} 
\hline
System & CPU Time (s) & Com. (MB) & Rounds  \\
\hline
DelphiDrowsy & 92.961 & 277.085 & - \\
CrypTFlowDrowsy & 0.254 & 87.496 & 59 \\
CryptoDrowsy & 0.027 & 5.271 & 0.411 \\
\hline
\end{tabular}\vspace{-0.2in}
\label{tab:sec_drowsy_comp}
\end{table}

The plaintext drowsiness detector (CompactCNN) achieves an accuracy of 0.8184 while the Delphi-based secure system performs slightly worse, at 0.7675. On the other hand, CrypTen and CrypTFlow2 achieve exactly the same results as the plaintext model, with accuracy scores of 0.8184 for each. The F1-Scores reflect this as well. CrypTen and CrypTFlow2 achieve the same results in plaintext at 0.8394, while Delphi's system results in an F1-score of 0.8064. 

This discrepancy may be related to the use of homomorphic encryption and numerical approximations in Delphi which help to achieve a higher level of security but reduces the accuracy of the system overall. In contrast, CrypTFlow2 and CrypTen use precise secret sharing schemes which enable more accurate computations. Moreover, CrypTFlow2 and CrypTen achieve better runtime performance by foregoing the expensive preprocessing computations required for homomorphic encryption schemes in favor of relatively cheap oblivious transfers in the case of CrypTflow2, and the use of a secure mediating agent in the case of CrypTen.

\subsection{Private Red-Light Running Detection}

We evaluate the performance of our privacy-preserving red light violation detection system by studying its underlying object detector in two ways. First, we compare the performance of our chosen private object detector, FastSec-YOLO, across a series of different image resolutions and batch sizes. For image and batch size, we report variations in the running time performance of the system. In addition, we analyze the detection accuracy and speed of our Fast-SecYOLO compared to alternative secure YOLOv5 systems. Second, we compare the efficiency of our chosen private system against the relevant secure benchmarks, including Faster-RCNN \cite{bi_achieving_2023,liu_privacy-preserving_2022} and YOLOv3-SPP \cite{zhou2022secure}. 

When running YOLOv5 in plaintext or secure settings, batch size and image size can be adjusted to balance inference speed and accuracy. In both settings, increasing the image size results in a monotonic increase in inference time, as shown in Figures \ref{graph:plain_img} and \ref{graph:sec_image}. However, larger image sizes also improve inference accuracy. Based on an empirical analysis of detection outcomes with FastSec-YOLO, an image size of $288\times 288$ was found to be optimal, offering the best trade-off between inference speed and accuracy for red light violation detection. Batch size, on the other hand, affects only inference speed. The optimal batch size is 16, as larger sizes slightly increase inference time. Performance does not degrade beyond a batch size of 32. This contrasts with plaintext YOLOv5, where inference time decreases monotonically with increasing batch size, stabilizing beyond a batch size of 32.

Table \ref{tab:fast_sec_yolo} summarizes the average CPU and GPU inference times, average precision for relevant RLR vehicle classes (1, 2, 3, 5, and 7), communication overhead, and communication rounds per image across batches of 32 images resized to $288 \times 288$. These results are based on the COCO128 baseline dataset, a subset of COCO \cite{lin2015microsoft}, used to train and evaluate the YOLOv5 object detector. FastSec-YOLO demonstrates significantly better runtime performance, requiring nearly 90 seconds less than CrypTen-YOLOv5x on the CPU and 25 seconds less on the GPU. However, this comes with a tradeoff: FastSec-YOLO, built on YOLOv5n—the smallest YOLOv5 model—shows notably lower average precision compared to CrypTen-YOLOv5x. This tradeoff may be acceptable for some instances, such as red light violation detection, where higher resolution vehicle classifications require only moderate precision. The plaintext RLR detection algorithm proposed by \cite{yahya2019fully}, based on YOLOv3, achieves an F1-score of 0.93 for identifying violators. Since FastSec-YOLO achieves precision comparable to YOLOv5s (equivalent to YOLOv3), our secure RLR detection system has similar performance as the plaintext version.

\vspace{-0.05in}

\begin{table}[!htbp]
\fontsize{9.5}{11}\selectfont
\centering
\caption{Average runtime, communication, and round complexity performance compared between our system, FastSec-YOLO and relevant benchmarks as reported.}
\vspace{-0.1in}
\small
\begin{tabular}{|c|c|c|c|} 
\hline
System & Model & Time (s) & Com. (MB) \\
\hline
FastSec-YOLO & YOLOv5n & \textbf{1.734} & 151.5 \\
SecRCNN \cite{liu_privacy-preserving_2022} & FasterRCNN & 180 & 150 \\
P2OD \cite{bi_achieving_2023} & FasterRCNN & 190.527 & 6166.723 \\
PPDF \cite{zhou2022secure} & YOLOv3-SPP & 276 & 163.1 \\
\hline
\end{tabular}\vspace{-0.1in}
\label{tab:secure_det_comparison}
\end{table}

\begin{table}[!htbp]
\centering
\caption{MAP50, Mean Precision (MP), Mean Recall (MR) and Runtime (efficiency) metrics for each of the models in a plaintext setting. Mean precision and recall account for all classes. Batch size: 32, and input resolution: 288 $\times$ 288.}
\vspace{-0.1in}
\small
\begin{tabular}[c]{|c|c|c|c|c|}
\hline 
Model & MAP50 & MP & MR & Time (s) \\
\hline
YOLOv5n & 0.4388 & 0.4939 & 0.3425 & 0.0027 \\
YOLOv5s & 0.4591 & 0.5425 & 0.3678 & 0.0047 \\
YOLOv5m & 0.6150 & 0.6584 & 0.5234 & 0.0091 \\
YOLOv5l & 0.6916 & 0.7618 & 0.6136 & 0.0172 \\
YOLOv5x & 0.7552 & 0.7837 & 0.6467 & 0.0332 \\
\hline
\end{tabular}
\label{tab:plain_val}
\end{table}

\begin{figure*}[!htbp]
    \centering
    \begin{subfigure}[b]{0.25\textwidth}
        \centering
        \includegraphics[width=\textwidth]{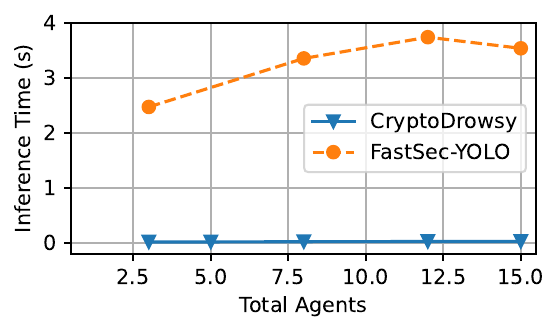}
        \caption{25\% Split Agent Types}
        \label{graph:025_agents}
    \end{subfigure}
    \hspace{-0.1in}
    \begin{subfigure}[b]{0.25\textwidth}
        \centering
        \includegraphics[width=\textwidth]{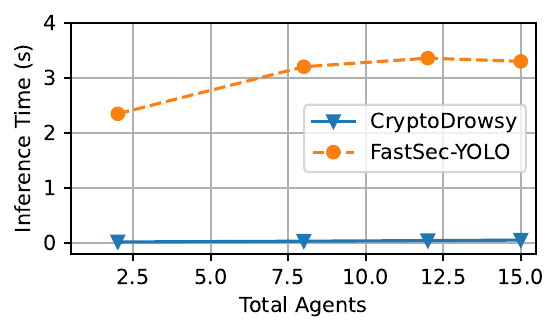}
        \caption{50\% Split Agent Types}
        \label{graph:05_agents}
    \end{subfigure}
    \hspace{-0.1in}
    \begin{subfigure}[b]{0.25\textwidth}
        \centering
        \includegraphics[width=\textwidth]{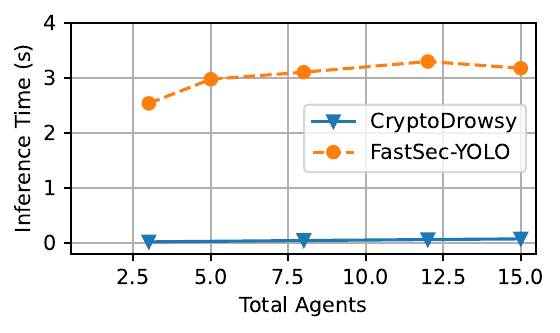}
        \caption{75\% Split Agent Types}
        \label{graph:075_agents}
    \end{subfigure}
    \hspace{-0.1in}
    \begin{subfigure}[b]{0.25\textwidth}
        \centering
        \includegraphics[width=\textwidth]{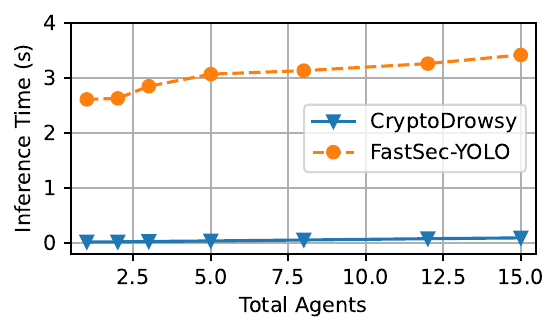}
        \caption{100\% Single Agent Type}
        \label{graph:100_agents}
    \end{subfigure}
       \caption{Evaluation of SecureV2X on Simultaneously Supporting Multiple Vehicles  (in different ratios). 
       Each subfigure displays the reported inference times when instances of CryptoDrowsy or FastSec-YOLO represent a given percentage of the systems (clients) being run. We conduct the evaluations by setting a total number of secure processes ranging from 1 to 15 at once. } 
    \label{fig:multi_agent}
\end{figure*}

Table \ref{tab:secure_det_comparison} compares the inference speed of FastSec-YOLO with other private object detectors as reported.\footnote{For prior works without open-sourced code, we included their reported results for comparison.}
This comparison demonstrates
the superior efficiency of our system. Whereas other secure systems all require
in excess of 180 seconds to compute (as reported), FastSec-YOLO requires only 1.734 seconds. This comes far closer to real-time performance than any of the prior work.

\subsection{Multi-Agent System Performance}

To evaluate how the efficiency of SecureV2X is affected by increased server loads while supporting multiple agents of different types, we conduct a series of multi-agent evaluations. In each setting, the server initializes a series of CryptoDrowsy and FastSec-YOLO processes simultaneously for a certain number of total clients. For our experiments, we set the number of total clients as either 1, 2, 3, 5, 8, 12, and 15 and the proportion of 
agents executing either CryptoDrowsy or FastSec-YOLO to 0.00, 0.25, 0.50, 0.75, or 1.00 while the remaining agents run the opposite protocol. Each agent running CryptoDrowsy processes the same set of 314 $1\times 384$ EEG vectors whereas each instance of FastSec-YOLO processes the same batch of 4 frames - scaled to a size of $288\times 288$ - from a real life red light violation recording.  

The results are shown in Figure \ref{fig:multi_agent}. Each subfigure illustrates the performance of CryptoDrowsy and FastSec-YOLO as their execution comprises a varying proportion of agents in the simulation. As expected, the average inference time per input increases with the number of agents. 
In all four sets of simulations, average 
inference times per agent are minimized where the total number of agents is the fewest. 
When more resources can be dedicated to a smaller set of agents, then 
their associated run times should be reduced. However, as the number of agents increases, 
the runtime increases only marginally. Specifically, the difference in average inference time per agent when the number of agents is 1 and 15 is at most 1 second for FastSec-YOLO (RLR Agents)
and 74 milliseconds for CryptoDrowsy (Driver Agents).

\section{Related Works}

In EEG-based drowsiness detection, Agarwal et al. \cite{agarwal_protecting_2019} proposes a secure linear regression system which makes use of secure multiparty-computation via additive secret sharing to jointly train linear regression models for predicting drowsiness states. As an alternate approach, Popsecu et al. \cite{popescu_privacy_2021} implement privacy preserving EEG signal classification through the use of homomorphic encryption. This approach preserves the correctness of linear operations over the data while maintaining its privacy. 

Several studies have explored secure implementations of object detection algorithms for applications in autonomous driving, traffic violation detection, and more. Due to the depth of object detection networks, homomorphic encryption proves inefficient for these computations. Instead, works such as \cite{zhou2022secure}, \cite{liu_privacy-preserving_2022}, and \cite{bi_achieving_2023} employ additive secret sharing with a trusted third party to distribute random multiplicative triples (Beaver's triples). However, these approaches are too slow for real-time applications, each requiring at least 180 seconds per inference. Conversely, Visor offers a more efficient but less secure solution \cite{rishabh2020visor}. This system uses a hybrid trusted execution environment (TEE) that optimizes CPU and GPU resources across vision tasks. For object detection, Visor securely implements the YOLO detector, achieving near-plaintext inference speeds. Despite its efficiency, TEEs are susceptible to side-channel attacks and other vulnerabilities that MPC can avoid.

\section{Conclusion}

In this work, we introduce SecureV2X, a unified framework for tackling safety- and privacy-critical challenges in V2X and intelligent transportation. We design CryptoDrowsy for fast, private drowsiness detection and FastSec-YOLO for real-time red light violation detection. SecureV2X significantly outperforms benchmark secure systems, with CryptoDrowsy achieving nearly $10\times$ faster inference than CrypTFlow2 and FastSec-YOLO exceeding the next fastest benchmark by over $100\times$. These efficiency gains enable practical, time-sensitive safety applications while preserving user data privacy and proprietary model security.

\section*{Acknowledgment}
This work is partially supported by the National Science Foundation (NSF) under Grants No. CNS-2302689, CNS-2308730, CNS-2319277, CNS-2432533, CMMI-2326341 and ITE-2452747, as well as by a Cisco Research Award.

\printbibliography

@article{cui_compact_2022,
	title = {A compact and interpretable convolutional neural network for cross-subject driver drowsiness detection from single-channel {EEG}},
	volume = {202},
	issn = {10462023},
	url = {https://linkinghub.elsevier.com/retrieve/pii/S1046202321001092},
	doi = {10.1016/j.ymeth.2021.04.017},
	language = {en},
	urldate = {2024-08-07},
	journal = {Methods},
	author = {Cui, Jian and Lan, Zirui and Liu, Yisi and Li, Ruilin and Li, Fan and Sourina, Olga and Müller-Wittig, Wolfgang},
	month = jun,
	year = {2022},
	pages = {173--184},
}

@article{thao_automatic_2022,
	title = {Automatic {Traffic} {Red}-{Light} {Violation} {Detection} {Using} {AI}},
	volume = {27},
	issn = {16331311, 21167125},
	url = {https://www.iieta.org/journals/isi/paper/10.18280/isi.270109},
	doi = {10.18280/isi.270109},
	language = {en},
	number = {1},
	urldate = {2024-08-07},
	journal = {Ingénierie des systèmes d information},
	author = {Thao, Le Quang and Cuong, Duong Duc and Anh, Nguyen Tuan and Anh, Pham Mai and Duc, Ha Minh and Minh, Nguyen},
	month = feb,
	year = {2022},
	pages = {75--80},
}

@article{ramzan2019survey,
    title={A survey on state-of-the-art drowsiness detection techniques},
    author={Ramzan, Muhammad and Khan, Hilkmat Ullah and Awan, Shahid Mahmood and Ismail, Amina and Ilyas, Mahwish and Mahmood Ahsan},
    journal={IEEE Access},
    year={2019},
    volume={7},
    pages={61904--61919},
    publisher={IEEE}
}

@article{lin2005eeg,
    title={EEG-based drowsiness estimation for safety driving using independent component analysis},
    author={Lin, Chin-Teng and Wu, Ruei-Cheng and Liang, Sheng-Fu and Chao, Wen-Hung and Chen, Yu-Jie and Jung, Tzyy-Ping},
    journal={IEEE Transactions on Circuits and Systems 1},
    year={2005},
    volume={52},
    number={12},
    pages={2726--2738},
    publisher={IEEE}
}

@misc{yahya2019fully,
    title={Fully-Automated-Red-Light-Violation-Detection},
    url={https://github.com/AhmadYahya97/Fully-Automated-red-light-Violation-Detection/tree/master},
    author={Yahya, Ahmad and Abdelkareem, Ihab and Al-fakhouri, Osama}
}

@inproceedings{kalal2010forward,
    author={Kalal, Zdenek and Mikolajczyk, Krystian and Matas, Jiri},
    title={{F}orward-{B}ackward {E}rror: {A}utomatic {D}etection of {T}racking {F}ailures},
    year={2010},
    booktitle={2010 20th International Conference on Pattern Recognition},
    publisher={IEEE},
    pages={2756--2759}
}

@inproceedings{bogdanov2008sharemind,
    author={Bogdanov, Dan and Laur, Sven and Willemson, Jan},
    title = {Sharemind: A Framework for Fast Privacy-preserving Computations},
    year={2008},
    booktitle={Computer Security - ESORICS 2008},
    publisher={Springer Berlin Heidelberg},
    pages={192--206}
}

@inproceedings{shah_novel_2023,
	title = {Novel {Crash} {Prevention} {Framework} for {C}-{V2X} using {Deep} {Learning}},
	url = {https://ieeexplore.ieee.org/document/10041397/?arnumber=10041397},
	doi = {10.1109/COMSNETS56262.2023.10041397},
	urldate = {2024-08-06},
	booktitle = {2023 15th {International} {Conference} on {COMmunication} {Systems} \& {NETworkS} ({COMSNETS})},
	author = {Shah, Foram N. and Patel, Dhaval K. and Shah, Kashish D. and Raval, Mehul S. and Zaveri, Mukesh and Merchant, S.N.},
	month = jan,
	year = {2023},
	note = {ISSN: 2155-2509},
	pages = {7--12},
}

@misc{lin2015microsoft,
      title={Microsoft COCO: Common Objects in Context},
      author={Tsung-Yi Lin and Michael Maire and Serge Belongie and Lubomir Bourdev and Ross Girshick and James Hays and Pietro Perona and Deva Ramanan and C. Lawrence Zitnick and Piotr Dollár},
      year={2015},
      eprint={1405.0312},
      archivePrefix={arXiv},
      primaryClass={cs.CV}
}

@article{liu_privacy-preserving_2022,
	title = {Privacy-{Preserving} {Object} {Detection} for {Medical} {Images} {With} {Faster} {R}-{CNN}},
	volume = {17},
	issn = {1556-6021},
	url = {https://ieeexplore.ieee.org/document/8864005/?arnumber=8864005},
	doi = {10.1109/TIFS.2019.2946476},
	urldate = {2024-08-05},
	journal = {IEEE Transactions on Information Forensics and Security},
	author = {Liu, Yang and Ma, Zhuo and Liu, Ximeng and Ma, Siqi and Ren, Kui},
	year = {2022},
	note = {Conference Name: IEEE Transactions on Information Forensics and Security},
	pages = {69--84},
}

@misc{knott_crypten_2022,
	title = {{CrypTen}: {Secure} {Multi}-{Party} {Computation} {Meets} {Machine} {Learning}},
	shorttitle = {{CrypTen}},
	url = {http://arxiv.org/abs/2109.00984},
	language = {en},
	urldate = {2024-08-05},
	publisher = {arXiv},
	author = {Knott, Brian and Venkataraman, Shobha and Hannun, Awni and Sengupta, Shubho and Ibrahim, Mark and van der Maaten, Laurens},
	month = sep,
	year = {2022},
	note = {arXiv:2109.00984 [cs]},
}

@article{agarwal_protecting_2019,
	title = {Protecting {Privacy} of {Users} in {Brain}-{Computer} {Interface} {Applications}},
	volume = {27},
	issn = {1558-0210},
	url = {https://ieeexplore.ieee.org/document/8755872/?arnumber=8755872},
	doi = {10.1109/TNSRE.2019.2926965},
	number = {8},
	urldate = {2024-08-05},
	journal = {IEEE Transactions on Neural Systems and Rehabilitation Engineering},
	author = {Agarwal, Anisha and Dowsley, Rafael and McKinney, Nicholas D. and Wu, Dongrui and Lin, Chin-Teng and De Cock, Martine and Nascimento, Anderson C. A.},
	month = aug,
	year = {2019},
	note = {Conference Name: IEEE Transactions on Neural Systems and Rehabilitation Engineering},
	pages = {1546--1555},
}

@article{bi_achieving_2023,
	title = {Achieving {Lightweight} and {Privacy}-{Preserving} {Object} {Detection} for {Connected} {Autonomous} {Vehicles}},
	volume = {10},
	issn = {2327-4662},
	url = {https://ieeexplore.ieee.org/document/9913215/?arnumber=9913215},
	doi = {10.1109/JIOT.2022.3212464},
	number = {3},
	urldate = {2024-08-05},
	journal = {IEEE Internet of Things Journal},
	author = {Bi, Renwan and Xiong, Jinbo and Tian, Youliang and Li, Qi and Choo, Kim-Kwang Raymond},
	month = feb,
	year = {2023},
	note = {Conference Name: IEEE Internet of Things Journal},
	pages = {2314--2329},
}

@inproceedings{khalfaoui_compare_2022,
    title = {{Comparative} {Study} of {YOLOv3} and {YOLOv5's} {Performances}
            for {Real-time} {Person} {Detection}},
    author = {Khalfaoui, Aicha and Badri, Abdelmajid and Mourabit, Ilham EL}, 
    booktitle = {2022 2nd International Conference on Innovative Research in Applied Science, 
                Engineering and Technology (IRASET)},
    year = {2022},
    pages = {1--5},
    publisher = {IEEE},
}

@article{nepal2022comparing,
    title={{C}omparing {YOLO}v3, {YOLO}v4, and {YOLO}v5 for {A}utonomous {L}anding
          {S}pot {D}etection in {F}aulty {UAV}s},
    author={Nepal, Upesh and Eslamiat, Hossein},
    year={2022},
    journal={Sensors},
    volume={22},
    number={2},
    pages={1--15},
    publisher={MDPI}
}

@inproceedings{damgard2012multiparty,
    title={Multiparty Computation from Somewhat Homomorphic Encryption},
    author={Damgård, Ivan and Pastro, Valerio and Smart, Nigel and Zakarias, Sarah},
    year={2012},
    booktitle={Proceedings of the 32nd Annual Cryptology Conference on Advances in Cryptology},
    pages={643--662},
    publisher={Springer Berlin Heidelberg}
}

@inproceedings{demmler2015aby,
    title={ABY - A Framework for Efficient Mixed-Protocol Secure Two-Party Computation},
    author={Demmler, Daniel and Schneider, Thomas and Zohner, Michael},
    year={2015},
    booktitle={NDSS Symposium 2015},
    pages={1--15},
    publisher={NDSS}
}

@inproceedings{goldreich1987how,
    title={How to Play any Mental Game, or a Completeness Theorem for Protocols with Honest Majority},
    author={Goldreich, Oded and Micali, Silvio and Widerson, Avi},
    year={1987},
    booktitle={Proceedings of the nineteenth annual ACM symposium on Theory of computing},
    pages={218--229},
    publisher={STOC}
}

@article{cohn2020red,
    title={Red Light Camera Interventions for Reducing Traffic Violations and 
          Traffic Crashes: A Systematic Review},
    author={Cohn, Ellen G. and Kakar, Suman and Perkins, Chloe and Steinbach, Rebecca and Edwards, Phil},
    journal={Campbell Systematic Reviews},
    volume={16},
    number={2},
    pages={1--52},
    year={2020}
}

@inproceedings{studdert2017once,
    title={Once Ticketed, Twice Shy? Specific Deterrence from Road Traffic Laws},
    author={Studdert, David M and Walter, Simon J and Goldhaber-Fiebert, Jeremy J},
    booktitle={Health Law Workshops},
    year={2017},
    publisher={Harvard Law School}
}

@article{zhang2023fine,
    title={A Fine-Grained Object Detection Model for Aerial Images based on YOLOv5 
          Deep Neural Network},
    author={Zhang, Rui and Xie, Cong and Deng, Liwei},
    journal={Chinese Journal of Electronics},
    volume={32},
    number={1},
    pages={51--63},
    year={2023}
}

@inproceedings{beaver1991efficient,
    title={Efficient Multiparty Protocols Using Circuit Randomization},
    author={Beaver, Donald},
    booktitle={11th Annual International Cryptology Conference},
    year={1991},
    pages={420--432},
    publisher={Springer}
}

@misc{aaa2019red,
    title={Red Light Running Crash Fatalities},
    author={{AAA}},
    year={2019},
    url={https://us.vocuspr.com/Newsroom/ViewAttachment.aspx?SiteName=AAACS&Entity=PRAsset&AttachmentType=F&EntityID=110661&AttachmentID=a05c7218-b5a8-4f9e-868d-557e5959f030}
}

@inproceedings{damgard2006unconditionally,
    title={Unconditionally Secure Constant-rounds Multi-party Computation for 
          Equality, Comparison, Bits and Exponentiation},
    author={Damgård, Ivan and Fitzi Matthias and Kiltz, Eike and Nielsen, Jesper Buus
           and Toft, Tomas},
    booktitle={Proceedings of the Third Conference on Theory of Cryptography},
    year={2006},
    pages={285--304},
    publisher={Springer Berlin Heidelberg}
}

@article{yusro_mod_comparison_2022,
    title = {{Comparison} of {Faster} {R-CNN} and {YOLOv5} for {Overlapping} {Objects} {Recognition}},
    author = {Yusro, Muhamad Munawarar and Ali, Rozniza and Hitam, Muhammad Suzuri},
    journal = {Baghdad Science Journal},
    year = {2023},
    volume = {20},
    number = {3}, 
    pages = {893--903}
}

@article{ahmed_pothole_detection_2021,
    title = {{Smart} {Pothole} {Detection} {Using} {Deep} {Learning} {Based} on
            {Dilated} {Convolution}},
    author = {Ahmed, Khaled R.},
    journal = {Sensors},
    year = {2021},
    volume = {21},
}

@article{popescu_privacy_2021,
	title = {Privacy {Preserving} {Classification} of {EEG} {Data} {Using} {Machine} {Learning} and {Homomorphic} {Encryption}},
	volume = {11},
	copyright = {http://creativecommons.org/licenses/by/3.0/},
	issn = {2076-3417},
	url = {https://www.mdpi.com/2076-3417/11/16/7360},
	doi = {10.3390/app11167360},
	language = {en},
	number = {16},
	urldate = {2024-08-05},
	journal = {Applied Sciences},
	author = {Popescu, Andreea Bianca and Taca, Ioana Antonia and Nita, Cosmin Ioan and Vizitiu, Anamaria and Demeter, Robert and Suciu, Constantin and Itu, Lucian Mihai},
	month = jan,
	year = {2021},
	note = {Number: 16
Publisher: Multidisciplinary Digital Publishing Institute},
	pages = {7360},
}

@inproceedings{zhou2022secure,
    author = {Zhou, Yongjie and Xiong, Jinbo and Bi, Renwan and Tian, Youliang},
    title = {{Secure} {YOLOv3-SPP}; {Edge-Cooperative} {Privacy-preserving} {Object} {Detection} for {Connected} {Autonomous} {Vehicles}},
    year = {2022},
    month = {December},
    booktitle = {2022 International Conference on Networking and Applications},
    publisher = {IEEE},
    pages = {82--89}
}

@inproceedings{rishabh2020visor,
    author = {Rishabh Poddar and Ganesh Ananthanarayanan and Srinath Setty and Stavros Volos and Raluca Ada Popa},
    title = {Visor: {Privacy-Preserving} Video Analytics as a Cloud Service},
    booktitle = {29th USENIX Security Symposium (USENIX Security 20)},
    year = {2020},
    month = {August},
    isbn = {978-1-939133-17-5},
    publisher = {USENIX Association},
    pages = {1039--1056},
    url = {https://www.usenix.org/conference/usenixsecurity20/presentation/poddar},
}

@inproceedings{cramer2005share,
  title={Share conversion, pseudorandom secret-sharing and applications to secure computation},
  author={Cramer, Ronald and Damg{\aa}rd, Ivan and Ishai, Yuval},
  booktitle={Theory of Cryptography: Second Theory of Cryptography Conference, TCC 2005, Cambridge, MA, USA, February 10-12, 2005. Proceedings 2},
  pages={342--362},
  year={2005},
  organization={Springer}
}

@misc{usdot_v2x,
    title={USDOT Releases National Deployment Plan for Vehicle-to-Everything (V2X) Technologies to Reduce Death and Serious Injuries on America’s Roadways},
    year = {2024},
    url={https://www.transportation.gov/briefing-room/usdot-releases-national-deployment-plan-vehicle-everything-v2x-technologies-reduce},
    author={US Department of Transportation}
}


\clearpage
\appendix 
\section{Sub-Protocols}\label{sec_subprotocols}


\begin{algorithm}[H]
    \caption{ConvBNSiLU, $\mathcal{F}_{ConvBNSiLU}$}
    \label{alg:convbnsilu}
    \begin{algorithmic}[1]
        \Statex \textbf{Input:} Parties $P_{k},k\in\{0,1\}$ hold input shares
        $\langle x\rangle_{k}$, weight and bias shares $\langle W_{i}\rangle_{k}$ and $\ashare{b_{i}}$ for $i\in\{0,1,2\}$
        \Statex \textbf{Output:} $P_{0}$ and $P_{1}$ respectively hold 
        arithmetic shares $\langle\mathcal{O}\rangle_{0}$ and $\langle\mathcal{O}\rangle_{1}$ of the output.
        \State $P_{k},k\in\{0,1\}$ compute: \\
        $\ashare{t_{0}}_{k} \leftarrow \mathcal{F}_{Conv2D}(\ashare{x}_{k},\ashare{W_{0}}_{k},\ashare{b_{0}}_{k})$ 
        \State $P_{k},k\in\{0,1\}$ compute: \\
        $\ashare{t_{1}}_{k}\leftarrow \mathcal{F}_{BatchNorm}(\ashare{t_{0}}_{k},\ashare{W_{1}}_{k},\ashare{b_{1}}_{k})$ 
        \State $P_{k},k\in\{0,1\}$ compute: $\ashare{\mathcal{O}}_{k}\leftarrow \mathcal{F}_{Swish}(\ashare{t_{1}}_{k})$
    \end{algorithmic}
\end{algorithm}


\begin{algorithm}[H]
    \caption{C3 Block, $\mathcal{F}_{C3}$}
    \label{alg:c3}
    \begin{algorithmic}[1]
        \Statex \textbf{Input:} Parties $P_{k},k\in\{0,1\}$ hold input shares
        $\langle x\rangle_{k}$, weight and bias shares $\langle W_{i}\rangle_{k}$ and $\ashare{b_{i}}_k$ for $i\in\{0,1,\dots,2n+2\}$
        \Statex \textbf{Output:} $P_{0}$ and $P_{1}$ respectively hold 
        arithmetic shares $\langle\mathcal{O}\rangle_{0}$ and $\langle\mathcal{O}\rangle_{1}$ of the output.
        \State $P_{k}, k\in\{0,1\}$ computes: \\ $\ashare{t_{0}}_{k}\leftarrow \mathcal F_{ConvBNSiLU}(\ashare{x}_{k},\ashare{W_{0}}_{k},\ashare{b_{0}}_{k})$ and $\ashare{t_{1}}_{k}\leftarrow \mathcal F_{ConvBNSiLU}(\ashare{x}_{k},\ashare{W_{1}}_{k},\ashare{b_{1}}_{k})$
        \For {$j=1,\dots,n$}
            \If {Bottleneck is type 1}
                \State $P_{k},k\in\{0,1\}$ computes: \\
                $\ashare{t_{0}}_{k} \leftarrow \mathcal{F}_{Bottleneck 1}(\ashare{t_{0}}_{k}, \ashare{W_{2j}}_{k}, \ashare{W_{2j+1}}_{k},$
                $\ashare{b_{2j}}_{k}, \ashare{b_{2j+1}}_{k})$
            \Else 
                \State $P_{k},k\in\{0,1\}$ computes: \\
                $\ashare{t_{0}}_{k} \leftarrow \mathcal{F}_{Bottleneck 2}(\ashare{t_{0}}_{k}, \ashare{W_{2j}}_{k}, \ashare{W_{2j+1}}_{k},$
                $\ashare{b_{2j}}_{k},\ashare{b_{2j+1}}_{k})$
            \EndIf
        \EndFor
        \State $P_{k},k\in\{0,1\}$ computes: \\
        $\ashare{t_{3}}_{k} \leftarrow \mathcal{F}_{Concat}(\ashare{t_{0}}_{k}, \ashare{t_{1}}_{k})$
        \State $P_{k},k\in\{0,1\}$ computes: \\
        $\ashare{\mathcal{O}}_{k}\leftarrow F_{ConvBNSiLU}(\ashare{t_{3}}_{k},\ashare{W_{2n+2}}_{k},\ashare{b_{2n+2}}_{k})$
    \end{algorithmic}
\end{algorithm}


\begin{algorithm}[H]
    \caption{Bottleneck 2, $F_{Bottleneck 2}$}\label{alg:bottleneck2}
    \begin{algorithmic}[1]
        \Statex \textbf{Input:} Parties $P_{k},k\in\{0,1\}$ hold input shares
        $\langle x\rangle_{k}$, weight and bias shares $\langle W_{i}\rangle_{k}$ and $\ashare{b_{i}}$ for $i\in\{0,1\}$
        \Statex \textbf{Output:} $P_{0}$ and $P_{1}$ respectively hold 
        arithmetic shares $\langle\mathcal{O}\rangle_{0}$ and $\langle\mathcal{O}\rangle_{1}$ of the output.
        \State $P_{k},k\in\{0,1\}$ computes: \\
        $\ashare{t_{0}}_{k} \leftarrow \mathcal{F}_{ConvBNSiLU}(\ashare{x}_{k},\ashare{W_{0}}_{k},\ashare{b_{0}}_{k})$
        \State $P_{k},k\in\{0,1\}$ computes: \\
        $\ashare{\mathcal{O}}_{k}\leftarrow\mathcal{F}_{ConvBNSiLU}(\ashare{t_{0}}_{k},\ashare{W_{1}}_{k},\ashare{b_{1}}_{k})$ 
    \end{algorithmic}
\end{algorithm}


\begin{algorithm}[H]
    \caption{Bottleneck 1, $F_{Bottleneck 1}$}
    \label{alg:bottleneck1}
    \begin{algorithmic}[1]
        \Statex \textbf{Input:} Parties $P_{k},k\in\{0,1\}$ hold input shares
        $\langle x\rangle_{k}$, weight and bias shares $\langle W_{i}\rangle_{k}$ and $\ashare{b_{i}}_k$ for $i\in\{0,1\}$
        \Statex \textbf{Output:} $P_{0}$ and $P_{1}$ respectively hold 
        arithmetic shares $\langle\mathcal{O}\rangle_{0}$ and $\langle\mathcal{O}\rangle_{1}$ of the output.
        \State $P_{k},k\in\{0,1\}$ computes: \\
        $\ashare{t_{0}}_{k} \leftarrow \mathcal{F}_{ConvBNSiLU}(\ashare{x}_{k},\ashare{W_{0}}_{k},\ashare{b_{0}}_{k})$
        \State $P_{k},k\in\{0,1\}$ computes: \\
        $\ashare{t_{1}}_{k}\leftarrow\mathcal{F}_{ConvBNSiLU}(\ashare{t_{0}}_{k},\ashare{W_{1}}_{k},\ashare{b_{1}}_{k})$ 
        \State $P_{k},k\in\{0,1\}$ computes: \\
        $\ashare{\mathcal{O}}_{k} \leftarrow \ashare{x}_{k} + \ashare{t_{1}}_{k}$
    \end{algorithmic}
\end{algorithm}

\begin{algorithm}[H]
    \caption{SPPF, $F_{SPPF}$}
    \label{alg:sppf}
    \begin{algorithmic}[1]
        \Statex \textbf{Input:} Parties $P_{k},k\in\{0,1\}$ hold input shares
        $\langle x\rangle_{k}$, weight and bias shares $\langle W_{i}\rangle_{k}$ and $\ashare{b_{i}}_{k}$ for $i \in \{0,1\}$
        \Statex \textbf{Output:} $P_{0}$ and $P_{1}$ respectively hold 
        arithmetic shares $\langle\mathcal{O}\rangle_{0}$ and $\langle\mathcal{O}\rangle_{1}$ of the output.
        \State $P_{k},k\in\{0,1\}$ computes: \\
        $\ashare{t_{0}}_{k} \leftarrow \mathcal{F}_{ConvBNSiLU}(\ashare{x}_{k},\ashare{W_{0}}_{k},\ashare{b_{0}}_{k})$
        \State $P_{k},k\in\{0,1\}$ computes: \\
        $\ashare{t_{1}}_{k} \leftarrow \mathcal{F}_{MaxPool2D}(\ashare{t_{0}}_{k})$ 
        \State $P_{k},k\in\{0,1\}$ computes: \\
        $\ashare{t_{2}}_{k} \leftarrow \mathcal{F}_{MaxPool2D}(\ashare{t_{1}}_{k})$
        \State $P_{k},k\in\{0,1\}$ computes: \\
        $\ashare{t_{3}}_{k} \leftarrow \mathcal{F}_{MaxPool2D}(\ashare{t_{2}}_{k}$
        \State $P_{k},k\in\{0,1\}$ computes: \\
        $\ashare{t_{4}}_{k} \leftarrow F_{Concat}(\ashare{t_{0}}_{k},\ashare{t_{1}}_{k}, \ashare{t_{2}}_{k}, \ashare{t_{3}}_{k})$ 
        \State $P_{k},k\in\{0,1\}$ computes: \\
        $\ashare{\mathcal{O}}_{k} \leftarrow F_{ConvBNSiLU}(\ashare{t_{4}}_{k}, \ashare{W_{1}}_{k}, \ashare{b_{1}}_{k})$
    \end{algorithmic}
\end{algorithm}

\end{document}